\documentclass[aps,twocolumn,showpacs,preprintnumbers,amsmath,amssymb]{revtex4}

\usepackage{graphicx}
\usepackage{dcolumn}
\usepackage{bm}
\usepackage{color}

\begin{document}


\title{Quantum phase slips in one-dimensional superfluids in a periodic potential}
\author{Ippei Danshita$^{1}$}
\author{Anatoli Polkovnikov$^{2}$}
\affiliation{
{$^1$Computational Condensed Matter Physics Laboratory, RIKEN, Wako, Saitama 351-0198, Japan}
\\
{$^2$Department of Physics, Boston University, Boston, MA 02215, USA}
}

\date{\today}

\begin{abstract}
We study the decay of superflow of a one-dimensional (1D) superfluid in the presence of a periodic potential. In 1D, superflow at zero temperature can decay via quantum nucleation of phase slips even when the flow velocity is much smaller than the critical velocity predicted by mean-field theories. Applying the instanton method to the $O(2)$ quantum rotor model, we calculate the nucleation rate of quantum phase slips $\Gamma$. When the flow momentum $p$ is small, we find that the nucleation rate per unit length increases algebraically with $p$ as $\Gamma/L \propto p^{2K-2}$, where $L$ is the system size and $K$ is the Tomonaga-Luttinger parameter. Based on the relation between the nucleation rate and the quantum superfluid-insulator transition, we present a unified explanation on the scaling formulae of the nucleation rate for periodic, disorder, and single-barrier potentials. Using the time-evolving block decimation method, we compute the exact quantum dynamics of the superflow decay in the 1D Bose-Hubbard model at unit filling. From the numerical analyses, we show that the scaling formula is valid for the case of the Bose-Hubbard model, which can quantitatively describe Bose gases in optical lattices.
\end{abstract}

\maketitle

\section{Introduction}
Recently, superfluidity and superconductivity in one dimension (1D) have been experimentally studied in various systems, including superconducting nanowires~\cite{giordano-88, bezryadin-00, altomare-06, li-11,arutyunov-08}, liquid helium in nanopores~\cite{toda-07,taniguchi-10}, and ultracold bosonic atoms in optical lattices~\cite{stoeferle-04, fertig-05, mun-07, haller-10}. A common property found in these different systems is that the transport in 1D is significantly suppressed compared to that in higher dimensions. This suppression of the transport might be interpreted as a consequence of stronger effects of thermal and quantum fluctuations in 1D. At temperatures higher than a certain characteristic value, thermal fluctuations allow the amplitude of the superfluid order parameter to vanish and its phase to unwind, leading to the decay of superflow~\cite{little-67, langer-67, mccumber-70}. Such a process is often referred to as a {\it phase slip}. When the temperature is sufficiently low, thermal fluctuations are suppressed and the nucleation of phase slips due to quantum tunneling provides dominant contributions to the superflow decay~\cite{zaikin-97, freire-97, nishida-99, kagan-00, buechler-01, cherny-11, khlebnikov-05, anatoli-05, schachenmayer-10}. Indeed some of the experiments with superconducting nanowires have observed the crossover from the regime of the thermal activation to the quantum regime~\cite{giordano-88, bezryadin-00, altomare-06, li-11,arutyunov-08}. 
{Moreover, the experiments studying the transport of 1D Bose gases in optical lattices~\cite{fertig-05} showed a good agreement with the theoretical analyses at zero temperature by the time-evolving block decimation (TEBD) method~\cite{danshita-09-1, montangero-09} that accurately takes into account the effects of quantum fluctuations. In light of this agreement, it is highly likely that the quantum regime has been achieved also in cold atom systems. Therefore, it is important to accurately calculate the nucleation rate of quantum phase slips for 1D superfluid.

In particular, the decay of 1D superflow in a periodic potential have attracted much interest because the transport of 1D atomic Bose gases has been studied in the presence of an optical lattice. In addition, in the experiments of liquid helium absorbed in nanopores, an inert layer of solid helium covering the wall of the pores acts as an external potential for 1D liquid helium~\cite{taniguchi-10}, which may be regarded as a periodic potential~\cite{eggel-11}. Having in mind ultracold atom experiments, most of previous theoretical studies have analyzed transport properties in the presence of a parabolic trapping potential in addition to periodic potentials by means of various numerical methods beyond mean-field approximations, such as exact diagonalization~\cite{ana-05}, truncated Wigner approximation~\cite{anatoli-04, ruostekoski-05}, fermionization method~\cite{rigol-05, pupillo-06}, and TEBD (or equivalently time-dependent density-matrix renormalization group)~\cite{danshita-09-1, montangero-09, anzi-11}. They have not explicitly related the 1D transport at zero temperature to quantum phase slips. In contrast, in Ref.~\cite{anatoli-05} the authors have pointed out the connection with phase slips and used the instanton method to analytically calculate the nucleation rate in a homogeneous lattice system. However, their instanton analyses in 1D have been restricted to the parameter regions far from the Mott transition and close to the superfluid critical velocity predicted by mean-field theory. 

In the present paper, we investigate quantum phase slips in 1D superfluids in the presence of a periodic potential. Applying the instanton techniques to the $O(2)$ quantum rotor model, we calculate the nucleation rate $\Gamma$ as a function of the flow (quasi-)momentum $p$. Since we treat the phase degrees of freedom on all the sites as independent variables in contrast to previous work that uses a single-collective-variable approximation~\cite{anatoli-05}, we can analyze the entire region of the momentum. Especially, when the momentum is much smaller than the critical value $p_{\rm c}$, we find that the nucleation rate per site obeys $\Gamma/L \propto (p/p_{\rm c})^{2K-2}$, where $L$ is the number of lattice sites and $K$ is the Tomonaga-Luttinger parameter. This power-law behavior with respect to the momentum means that the lifetime of superflow can be practically infinitely long if $p \ll p_{\rm c}$, namely the presence of superfluidity in 1D. It should be noted that a similar power-law behavior of the nucleation rate has been previously found in the case of 1D superfluid in the presence of a single-barrier potential~\cite{kagan-00, buechler-01} or a disorder potential~\cite{khlebnikov-05}, i.e. $\Gamma \propto (p/p_{\rm c})^{2K-1}$ for a single barrier and $\Gamma/L \propto (p/p_{\rm c})^{2K-1}$ in the case of disorder.  We emphasize that in the case of periodic potentials the exponent takes a different value. We discuss this difference from a viewpoint of the relation between the nucleation rate and the quantum superfluid-insulator transition. Moreover, we investigate the real-time dynamics of the superflow decay via quantum phase slips in the 1D Bose-Hubbard model (BHM) at unit filling by means of the quasi-exact numerical method of TEBD~\cite{vidal-04, danshita-09-2}. From the TEBD calculations, we confirm the validity of the scaling formula in the model that is quantitatively relevant to experiments with ultracold Bose gases confined in optical lattices.

The remainder of the paper is organized as follows. In Sec.~\ref{sec:QR}, we introduce BHM describing 1D bosons in a periodic potential and briefly review a derivation of the $O(2)$ quantum rotor model from BHM. In Sec.~\ref{sec:ins}, we calculate the nucleation rate of quantum phase slips using the instanton techniques and find a scaling formula of the nucleation rate with respect to the momentum. In Sec.~\ref{sec:dga}, we discuss a physical interpretation of the scaling formula from a viewpoint of the relation between the phase-slip nucleation and the superfluid-insulator transition. In Sec.~\ref{sec:TEBD}, applying TEBD to BHM at unit filling, we compute the quantum dynamics of superflow, which exponentially decays in time. We numerically extract the nucleation rate of quantum phase slips and compare it with the scaling formula. In Sec.~\ref{sec:sum}, we summarize our results and describe possible future work.
 
\section{$O(2)$ quantum rotor model}
\label{sec:QR}
We consider a system of 1D lattice bosons in a ring-shaped geometry, i.e. with a periodic boundary. To study such a system, one often uses BHM that especially allows for a quantitative description of Bose gases confined in optical lattices. The quantitative validity of BHM has been confirmed for both equilibrium and dynamical properties through the thorough comparisons between experiments on Bose gases in optical lattices and exact numerical analyses on the Bose-Hubbard model~\cite{trotzky-08, trotzky-10, trotzky-11, kasztelan-11}. However, before directly analyzing BHM, in order to gain analytical insights we use the $O(2)$ quantum rotor model that can be derived from BHM under a certain condition. The formulation of the instanton method is much simpler for the quantum rotor model~\cite{anatoli-05, danshita-10}. In this section, we review a derivation of the quantum rotor model.

We start with the grand canonical partition function,
\begin{eqnarray}
Z=\int {\cal D}b^{\ast}{\cal D}b \exp\left\{-\frac{S[b^{\ast},b]}{\hbar}\right\},
\label{part_func}
\end{eqnarray}
where the action $S[b^{\ast},b]$ for the BHM is given by
\begin{eqnarray}
S[b^{\ast},b] = \sum_{j=1}^L \int_{-\frac{\hbar\beta}{2}}^{\frac{\hbar\beta}{2}} d\tau   
\nonumber \\
&&\!\!\!\!\!\!\!\!\!\!\!\!\!\!\!\!\!\!\!\!\!\!\!\!\!\!\!\!\!\!\!\!\!\!\!\!\!\!\!\!\!\!\!\!\!\!\!\!\!\!\!\!\!\!\!
\times \left[
b_j^{\ast}(\tau)\hbar\frac{\partial}{\partial\tau}b_j(\tau)
- J\left( b_{j}^{\ast}(\tau)b_{j+1}(\tau) + b_{j+1}^{\ast}(\tau)b_{j}(\tau)\right)
\right.
\nonumber\\
&& \!\!\!\!\!\!\!\!\!\!\!\!\!\!\!\!\!\!\!\!\!\!\!\!\!\!\!\!\!\!\!\!\!\!\!\!\!\!\!\!\!\!\!\!\!\!\! \left.
+ \frac{U}{2}b_j^{\ast}(\tau)b_j^{\ast}(\tau)b_{j}(\tau)b_{j}(\tau)-\mu b_j^{\ast}(\tau) b_j(\tau)
\right],
\end{eqnarray}
where $U$ is the onsite interaction, $J$ the hopping energy, and $L$ the number of lattice sites.
Here, $\mu\approx U\nu$ is the chemical potential and $\nu$ is the filling factor. For convenience we introduce finite small temperature $T$ corresponding to the inverse temperature $\beta \equiv (k_B T)^{-1}$. In the end of calculations we will take the limit of $T\to 0$. Inserting $b_j=\sqrt{n_j}e^{i\theta_j}$, the action is rewritten as
\begin{eqnarray}
S[n,\theta]&=&\sum_{j=1}^L \int_{-\frac{\hbar\beta}{2}}^{\frac{\hbar\beta}{2}} d\tau
\left[
\hbar n_j\left( i \frac{\partial \theta_j}{\partial \tau} +\frac{1}{2n_j}\frac{\partial n_j}{\partial \tau}\right)
\right.
\nonumber \\
&&\!\!\!\!\!\!\!\!\!\!\!\!\!\!\!
\left. -2\sqrt{n_j n_{j+1}}J\cos\left(\theta_{j+1}-\theta_j\right)
+\frac{U}{2}(n_j-\nu)^2
\right].
\end{eqnarray}
We split the number of particles per site into its average and fluctuation as $n_j = \nu + \delta n_j$, and assume that $U\nu \gg J$ and $\nu \gg \delta n_j$.
Then, we find that the action is approximated as
\begin{eqnarray}
S[n,\theta] &=& \sum_{j=1}^L \int_{-\frac{\hbar\beta}{2}}^{\frac{\hbar\beta}{2}} d\tau
\left[
i\hbar\, (\nu+\delta n_j) \frac{\partial \theta_j}{\partial \tau} \right.
\nonumber \\
&& \,\,\,\,\,\,
\left. -2\nu J\cos\left(\theta_{j+1}-\theta_j\right)
+\frac{U}{2}\delta n_j^2
\right].
\label{eq:actionNphi}
\end{eqnarray}
Since Eq.~(\ref{eq:actionNphi}) contains only the linear and quadratic terms with respect to number fluctuations $\delta n_j$, these degrees of freedom can be integrated out. 
Then, the action is described in terms of the phases as
\begin{eqnarray}
S[\theta]&=&\sum_{j=1}^L \int_{-\frac{\hbar\beta}{2}}^{\frac{\hbar\beta}{2}} d\tau
\left[
i\hbar \nu \frac{\partial \theta_j}{\partial \tau}
+ \frac{\hbar^2}{2U}  \left(\frac{\partial \theta_j}{\partial \tau}\right)^2 \right.
\nonumber \\
&& \,\,\,\,\,\,\, \left.
-2\nu J\cos\left(\theta_{j+1}-\theta_j\right)
\right].
\label{eq:ac_phi_0}
\end{eqnarray}
When the filling factor is irrational, the first term makes the net contribution of the instanton or bounce solutions to the partition function to be zero~\cite{kashurnikov-96}. 
When the filling factor is integer (commensurate filling), the first term is necessarily equal to $\hbar \times 2\pi l$, where $l$ is an integer, and its contribution to the partition function is unity regardless of the trajectory of $\theta_j$.
In the latter case, the effective action takes the form of the quantum rotor model,
\begin{eqnarray}
S[\theta]\!=\!\sum_{j=1}^L \! \int_{-\frac{\hbar\beta}{2}}^{\frac{\hbar\beta}{2}} d\tau \!\!
\left[
\frac{\hbar^2}{2U}  \left(\frac{\partial \theta_j}{\partial \tau}\right)^2
\!\!\! -2\nu J\cos\left(\theta_{j+1}-\theta_j\right)
\right].
\label{eq:ac_phi_1}
\end{eqnarray}
Introducing the dimensionless parameter $K \equiv \pi \sqrt{2 \nu J / U}$ and the sound velocity $u\equiv d\sqrt{2\nu J U}/\hbar$, the action can be rewritten as
\begin{eqnarray}
S[\theta]\!=\!\frac{\hbar K}{2\pi} \! \sum_{j=1}^L \! \int_{-\frac{\hbar\beta}{2}}^{\frac{\hbar\beta}{2}} 
\! d\tau \!\!
\left[
\frac{d}{u}  \left(\frac{\partial \theta_j}{\partial \tau}\right)^2
\!\!\! -2\frac{u}{d}\cos\left(\theta_{j+1} \! - \! \theta_j\right)
\right].
\label{eq:ac_phi_2}
\end{eqnarray}
If one takes the continuum limit of $d\rightarrow 0$ and $L\rightarrow \infty$ while fixing the value of $Ld$, the action of Eq.~(\ref{eq:ac_phi_2}) coincides with that for the spinless Tomonaga-Luttinger (TL) liquid~\cite{giamarchi-04, cazalilla-11},
\begin{eqnarray}
S[\theta]=\frac{\hbar K}{2\pi}\int_{0}^{Ld} dx \int_{-\frac{\hbar\beta}{2}}^{\frac{\hbar\beta}{2}} d\tau
\left[
\frac{1}{u}  \left(\frac{\partial \theta}{\partial \tau}\right)^2
+u\left(\frac{\partial \theta}{\partial x}\right)^2
\right].
\label{eq:ac_TLL}
\end{eqnarray}
where it is obvious that $K$ is the TL parameter. Notice, however, that the values of $K$ and $u$ in Eq.~(\ref{eq:ac_TLL}) are renormalized due to the effects of high-energy modes and Umklapp scattering and that the original relations of those parameters with $U/J$ and $\nu$ no longer hold. The TL liquid model generally describes low-energy physics of a massless one-dimensional system. This indicates that while we analyze the discrete quantum rotor model of Eq.~(\ref{eq:ac_phi_2}), low-energy properties found in our analyses should be general in the spinless TL liquid.

It is convenient to express the imaginary time in units of the Josephson plasma time $\hbar/E_{\rm J}$ as
\begin{eqnarray}
\tau = \frac{\hbar}{E_{\rm J}} \tilde{\tau},
\label{eq:time}
\end{eqnarray}
where $E_{\rm J}\equiv \hbar u/(\sqrt{2}d) $ is the Josephson plasma energy.
Inserting Eq.~(\ref{eq:time}) into Eq.~(\ref{eq:ac_phi_2}), we obtain
\begin{eqnarray}
S=\hbar \frac{K}{\sqrt{2}\pi}\tilde{s},
\label{eq:rotorS}
\end{eqnarray}
where $\tilde{s}$ is the dimensionless action
\begin{eqnarray}
\tilde{s} = \int_{-\frac{\tilde{\beta}}{2}}^{\frac{\tilde{\beta}}{2}}
d\tilde{\tau} \left[
\frac{1}{2}\frac{\partial \vec{\theta}}{\partial \tilde{\tau}}\cdot \frac{\partial \vec{\theta}}{\partial \tilde{\tau}}
+V(\vec{\theta})
\right].
\label{eq:sVec}
\end{eqnarray}
In order to express the action compactly, we introduced in Eq.~(\ref{eq:sVec}) an $L$-dimensional vector $\vec{\theta}$  defined as
\begin{eqnarray}
\vec{\theta} =
\left(
\theta_{1}(\tilde{\tau}),\ldots, \theta_{j}(\tilde{\tau}), \ldots, \theta_{L}(\tilde{\tau})
\right)^{\bf t}
\end{eqnarray}
and the potential,
\begin{eqnarray}
V(\vec{\theta}) &=& \sum_{j=1}^{L} V_j(\theta_{j+1},\theta_j) \nonumber\\
&=& \sum_{j=1}^{L}  -2\cos\left( \theta_{j+1} - \theta_j \right).
\end{eqnarray}
%
%
%
and $\tilde{\beta} = \beta E_{\rm J}$. From Eqs.~(\ref{part_func}) and (\ref{eq:rotorS}) we clearly see that $h_{\rm e}\equiv \sqrt{2}\pi/K$ plays the role of the effective dimensionless Planck's constant for this problem.  The limit of $h_{\rm e}\to 0$ corresponds to the classical (Gross-Pitaevskii) regime, while at $h_{\rm e}\gtrsim 1$ quantum fluctuations become significant and can even drive the system to a different insulating phase.
 
\section{Quantum nucleation rate for the phase slips}
\label{sec:ins}
Extremizing the action by imposing $\delta \tilde{s} = 0$, we obtain the classical equations of motion for the phases $\theta_j$,
\begin{eqnarray}
\frac{\partial^2 \theta_j}{\partial \tilde{\tau}^2} =
-2\sin\left(\theta_{j+1}-\theta_j \right)
+ 2\sin\left(\theta_{j}-\theta_{j-1} \right).
\label{eq:classical}
\end{eqnarray}
There are two types of stationary solution of Eq.~(\ref{eq:classical}).
The first one describing a state carrying a homogeneous superflow with the winding-number $n$ is
\begin{eqnarray}
\theta^M_{n, j} = \frac{2\pi n}{L}(j-1),
\label{eq:meta}
\end{eqnarray}
This state possesses the (quasi-)momentum $p=2\pi\hbar n/(Ld)$. The other is a saddle-point solution with a phase kink separating two states with different winding numbers:
\begin{eqnarray}
\theta^S_{n, j} =\frac{\alpha_n}{2}+\varphi_n(j-1),
\label{eq:saddle}
\end{eqnarray}
where
\begin{eqnarray}
\alpha_n = - \pi\frac{L-1+2n}{L-2} \mod 2\pi,
\end{eqnarray}
and
\begin{eqnarray}
\varphi_n = \frac{2\pi n-\alpha_n}{L-1}.
\end{eqnarray}
The value of $\varphi_n$ defines the momentum in the system and $\alpha_n$ is the phase difference between the 1st and $L$-th sites, the location of the phase kink. The magnitude of this kink $\alpha_n$ is defined within the interval $[-2\pi,0]$. In the limit of the large number of sites $L\gg n$ the expression for $\alpha_n$ simplifies:
\begin{equation}
\alpha_n\approx -\pi\left(1+{1+ 2 n\over L}\right) \mod 2\pi.
\end{equation}
For small winding numbers $n\ll L$ the phase kink is approximately equal to $\pi$.

%
%
%
%
%
%

We consider that a metastable flowing state with the winding number $n$ is prepared at the initial time $t=0$. The flow momentum is assumed to be smaller than the critical value $p_{\rm c}=\hbar\pi /(2d)$, above which the uniform flowing solutions become unstable. In the classical limit ($h_{\rm e} \rightarrow 0$) and at zero temperature the system remains in the initial state for an infinitely long time, i.e. the superflow is persistent. In contrast, when quantum fluctuations are strong enough, the metastable state decays into states with smaller momenta through quantum nucleation of phase slips and the lifetime of the superflow is finite. The decay rate of the metastable state, i.e. the nucleation rate of the quantum phase slip, can be calculated using the celebrated instanton formula~\cite{coleman-77,callan-77,polyakov-77,coleman-85,sakita-85}: 
\begin{eqnarray}
\hbar \Gamma \simeq E_{\rm J} L A \sqrt{\frac{\tilde{s}_B}{2\pi h_{\rm e}}}
\exp\left(-\frac{\tilde{s}_B}{h_{\rm e}}\right).
\label{eq:decayGam}
\end{eqnarray}
Here $\tilde{s}_B$ is the action for the bounce solution $\vec{\theta}(\tilde{\tau})=\vec{\theta}^B(\tilde{\tau})$ of Eq.~(\ref{eq:classical}), and the prefactor $A$ is given by
\begin{eqnarray}
A=\left|\frac{\prod_{m}\lambda_m^{(0)}}{\prod_{m \neq 0} \lambda_m}\right|^{1/2},
\label{eq:A}
\end{eqnarray}
where $\lambda_m$'s and $\lambda_m^{(0)}$'s are the solutions of the following eigenvalue equations:
\begin{eqnarray}
\hat{{\cal M}}\vec{\xi}_m(\tilde{\tau})=\lambda_m\vec{\xi}_m(\tilde{\tau}),
\label{eq:eigEqM1}
\end{eqnarray}
and
\begin{eqnarray}
\hat{{\cal M}}^{(0)}\vec{\xi}_m^{(0)}(\tilde{\tau})=\lambda_m^{(0)}\vec{\xi}_m^{(0)}(\tilde{\tau}).
\label{eq:eigEqM2}
\end{eqnarray}
Here the $L$-dimensional vectors
\begin{eqnarray}
\vec{\xi}_m =
\left(
\xi_{1,m}(\tilde{\tau}),\ldots, \xi_{j,m}(\tilde{\tau}), \ldots, \xi_{L,m}(\tilde{\tau})
\right)^{\bf t},
\end{eqnarray}
and
\begin{eqnarray}
\vec{\xi}_m^{(0)} =
\left(
\xi_{1,m}^{(0)}(\tilde{\tau}),\ldots, \xi_{j,m}^{(0)}(\tilde{\tau}), \ldots, \xi_{L,m}^{(0)}(\tilde{\tau})
\right)^{\bf t}.
\end{eqnarray}
obey the orthonormalization condition
\begin{eqnarray}
\int d\tilde{\tau} \,\, \vec{\xi}_l \cdot \vec{\xi}_m = \delta_{l,m},
\,\,\,
\int d\tilde{\tau} \,\, \vec{\xi}_l^{(0)} \cdot \vec{\xi}_m^{(0)} = \delta_{l,m}.
\end{eqnarray}
The $L\times L$ matrices $\hat{{\cal M}}$ and $\hat{{\cal M}}^{(0)}$ are defined as
\begin{eqnarray}
{\cal M}_{j,k}&&\!\!\!\!=\delta_{j,k}
\left(
-\frac{\partial^2}{\partial \tilde{\tau}^2}
+ \left. \frac{\partial^2 V_j}{\partial \theta_j^2} \right|_{\vec{\theta}=\vec{\theta}^B}
+ \left. \frac{\partial^2 V_{j-1}}{\partial \theta_j^2} \right|_{\vec{\theta}=\vec{\theta}^B}
\right)
\nonumber\\
&&\!\!\!\!\!\!\!\!\!\!\!\!\!\!\!\!\!\!\!\!\! + \delta_{j, k-1}
\left. \frac{\partial^2 V_j}{\partial \theta_j \partial \theta_{j+1}} \right|_{\vec{\theta}=\vec{\theta}^B}
+ \delta_{j, k+1}
\left. \frac{\partial^2 V_{j-1}}{\partial \theta_j \partial \theta_{j-1}} \right|_{\vec{\theta}=\vec{\theta}^B},
\end{eqnarray}
and
\begin{eqnarray}
{\cal M}_{j,k}^{(0)}=\delta_{j,k}
\left(
-\frac{\partial^2}{\partial \tilde{\tau}^2}
+2 \omega^2
\right)
- \delta_{j, k-1} \omega^2
- \delta_{j, k+1} \omega^2,
\end{eqnarray}
where $\omega^2= \left. \partial_{\theta_j}^2 V_j \right|_{\vec{\theta}=\vec{\theta}_n^M}$. 
Note that the factor of $L$ in the right hand side of Eq.~(\ref{eq:decayGam}) reflects the fact that there are $L$ independent trajectories corresponding to the phase slip happening at one out of $L$ links~\cite{danshita-10, callan-77}. We emphasize that the use of the quantum rotor model is advantageous in the sense that $\tilde{s}_{\rm B}$ and $A$ do not depend on $\nu$ and $U/J$ but depend only on $p$ and $L$. This means that $\hbar\Gamma/E_{\rm J}$ depends on $U$, $J$, and $\nu$ only through $h_{\rm e}$.

\begin{figure}[htb]
\includegraphics[scale=0.5]{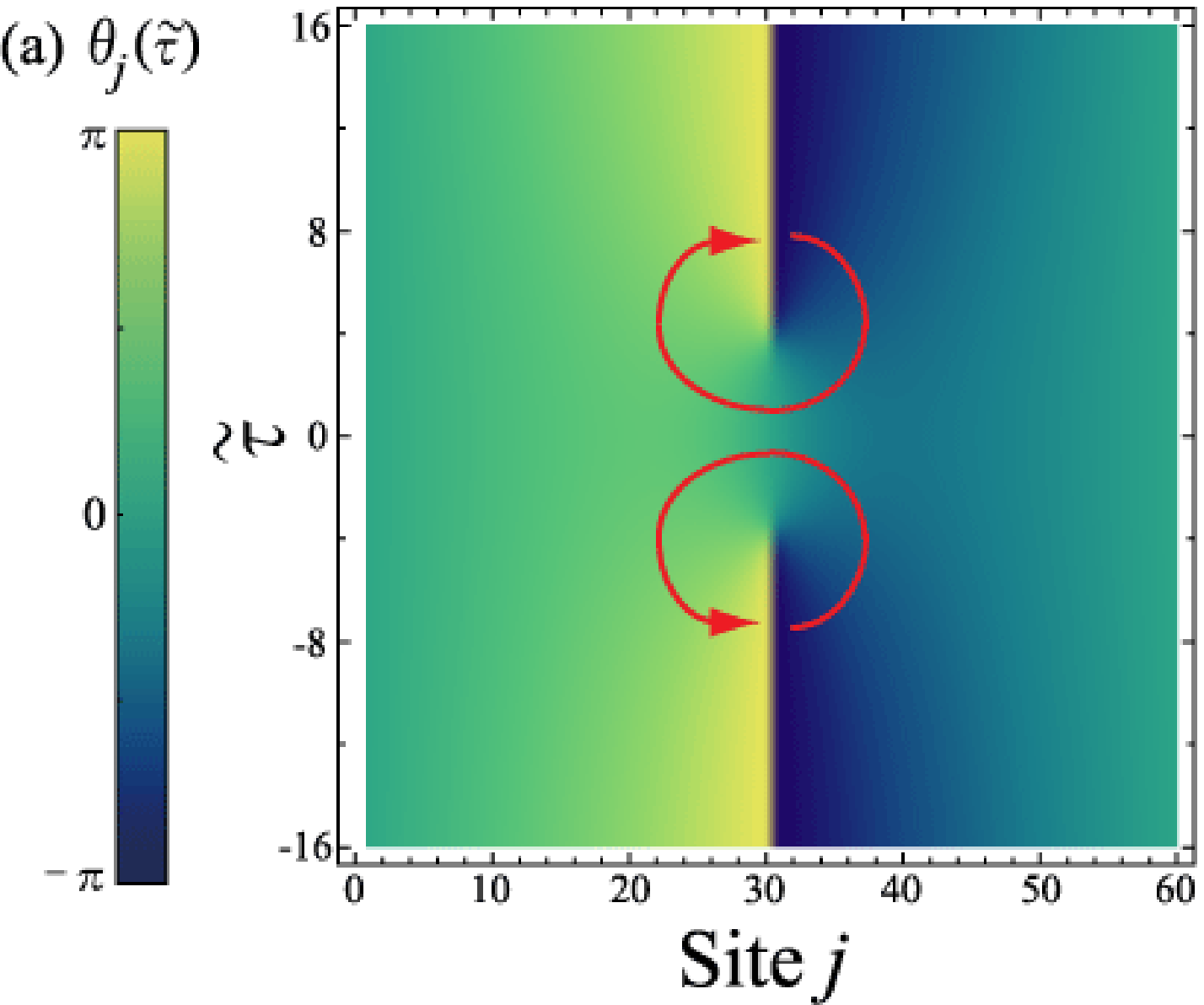}
\includegraphics[scale=0.6]{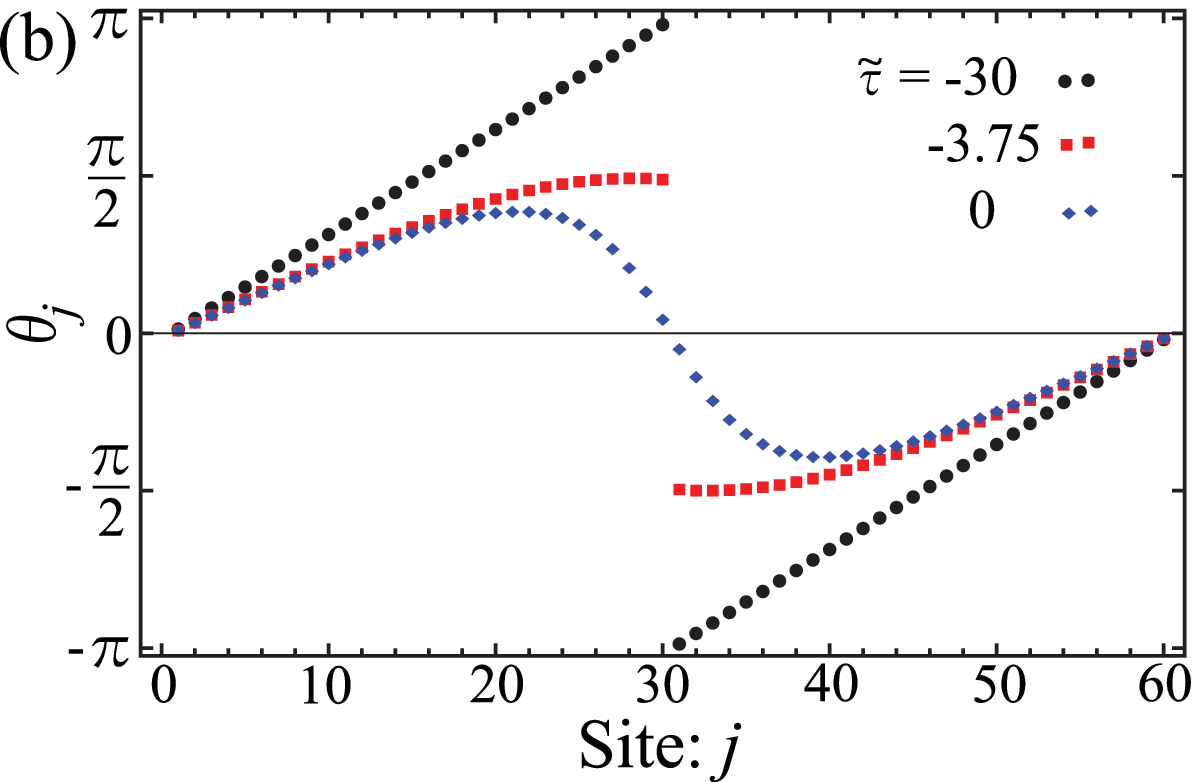}
\includegraphics[scale=0.65]{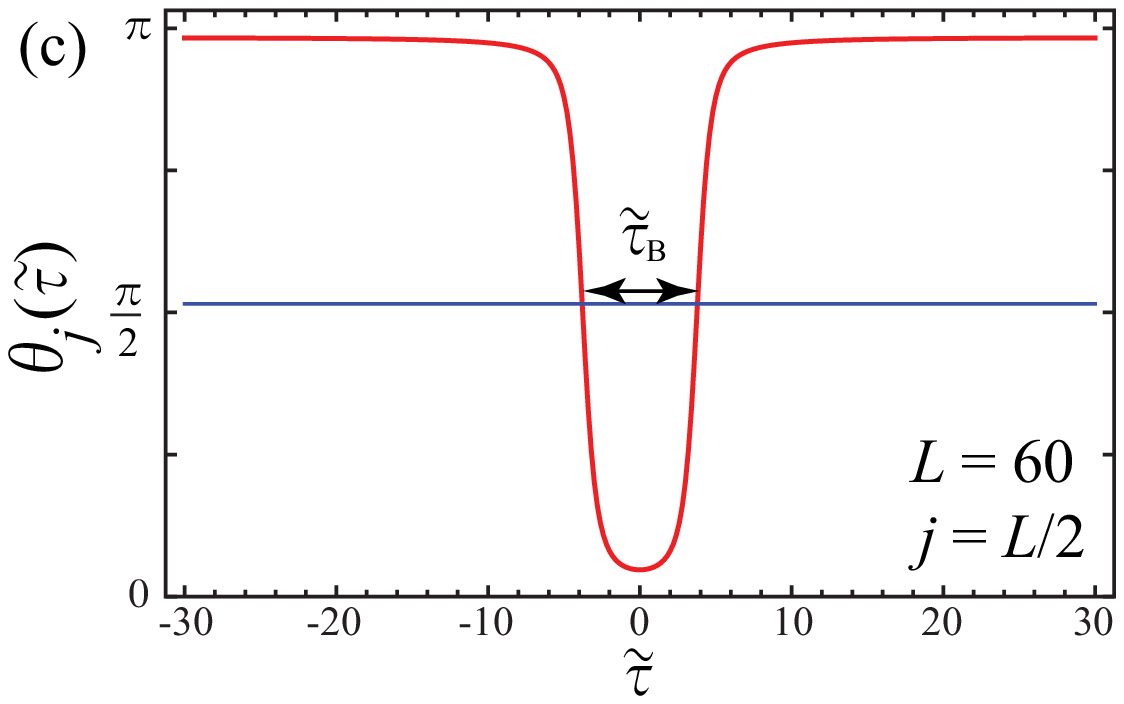}
\caption{\label{fig:bounce}
(Color online) (a) The bounce solution for the initial state with the winding number $n=1$. We take $L=60$. It is clear that the bounce solution forms a pair of a vortex and an anti-vortex in the $(x,\tau)$-plane. 
(b) The bounce solution is shown in section along the lines $\tilde{\tau}=-30$ (black circles), $-3.75$ (red squares), and $0$ (blue diamonds), which correspond to the metastable state, the vicinity of the saddle point, and the classical turning point.
(c) The bounce solution is shown in section along the line $j=L/2$.
}
\end{figure}
\begin{figure}[tb]
\includegraphics[scale=0.6]{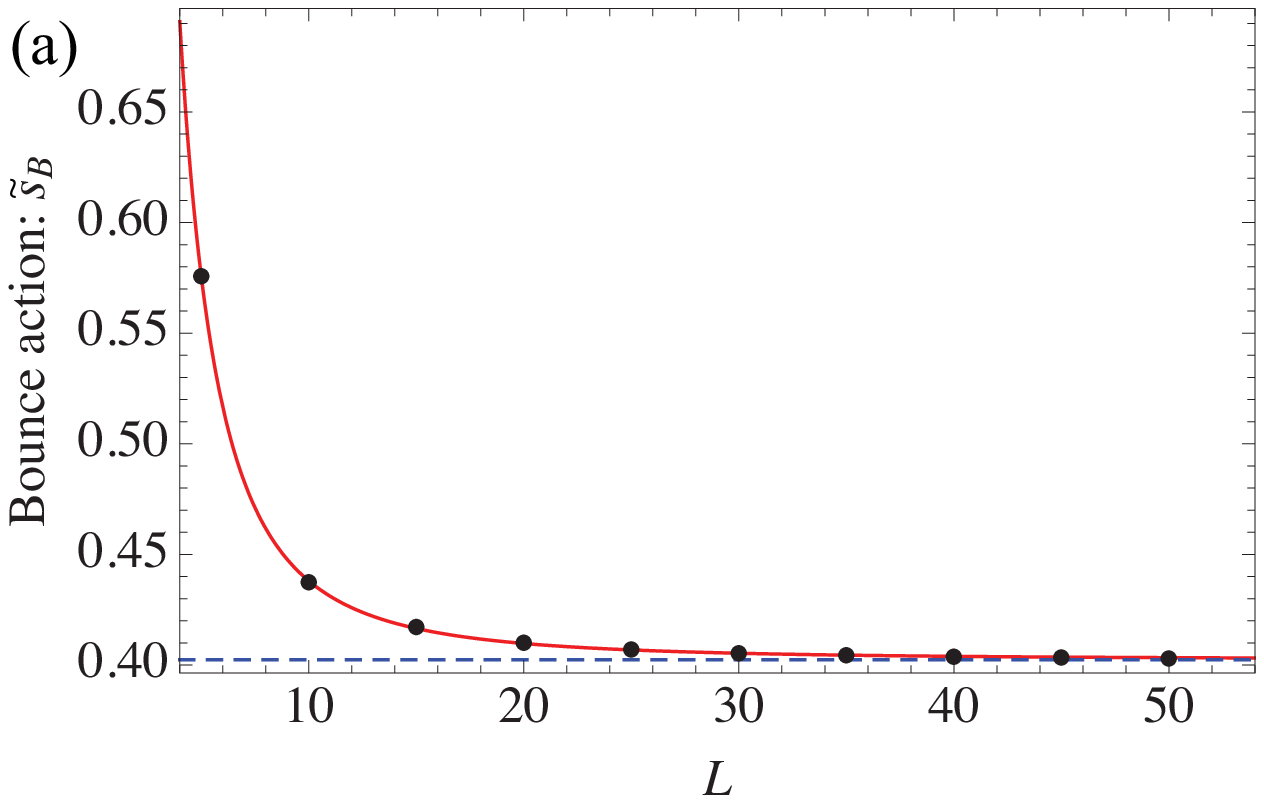}
\includegraphics[scale=0.6]{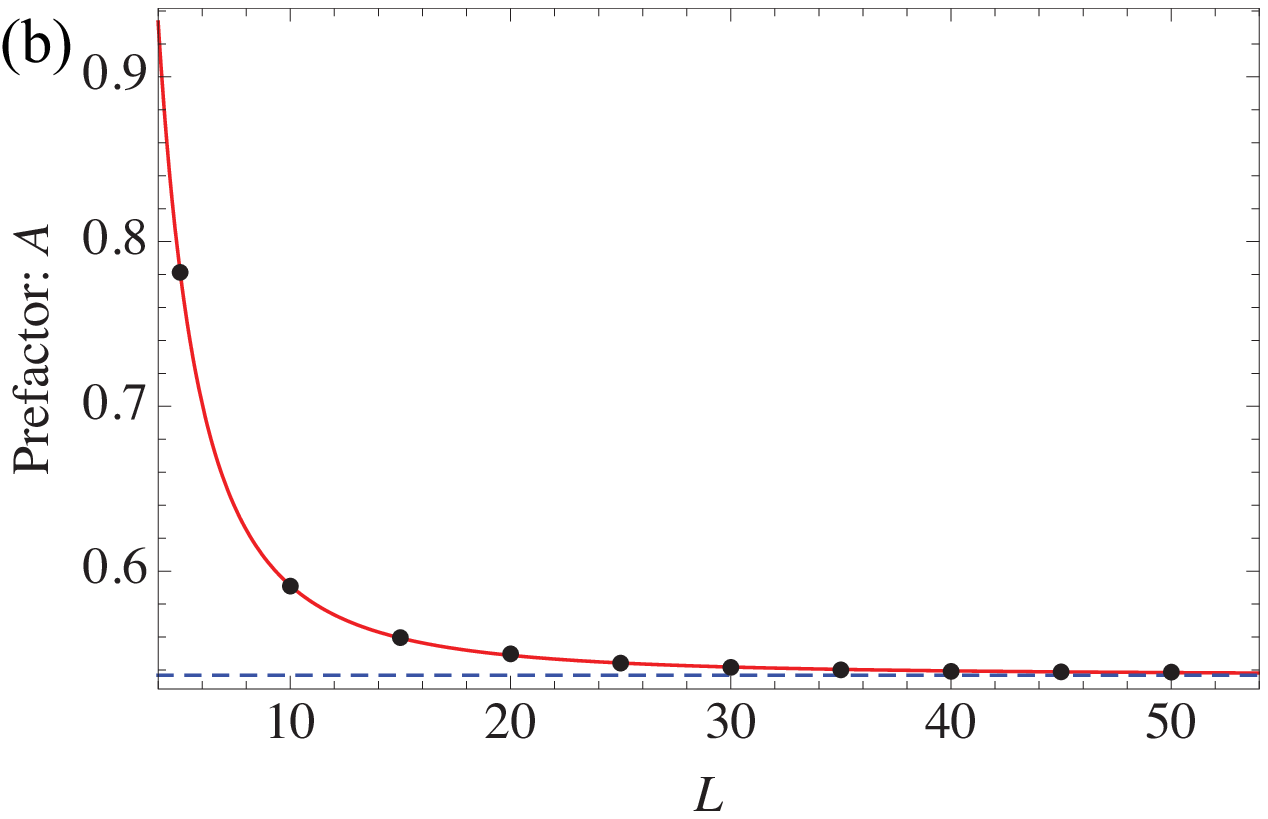}
\caption{\label{fig:ConvSBA}
(Color online) (a) The bounce action $\tilde{s}_B$ versus the system size $L$.
(b) The prefactor $A$ versus the system size $L$.
The (quasi-)momentum is fixed to be $pd/\hbar = 2\pi/5$.
The red solid line represents the best fit to the data with a fitting function $f(x)=a+bx^{-c}$.
The blue dashed line represents the extrapolated value $a$.
}
\end{figure}
\begin{figure}[htb]
\includegraphics[scale=0.6]{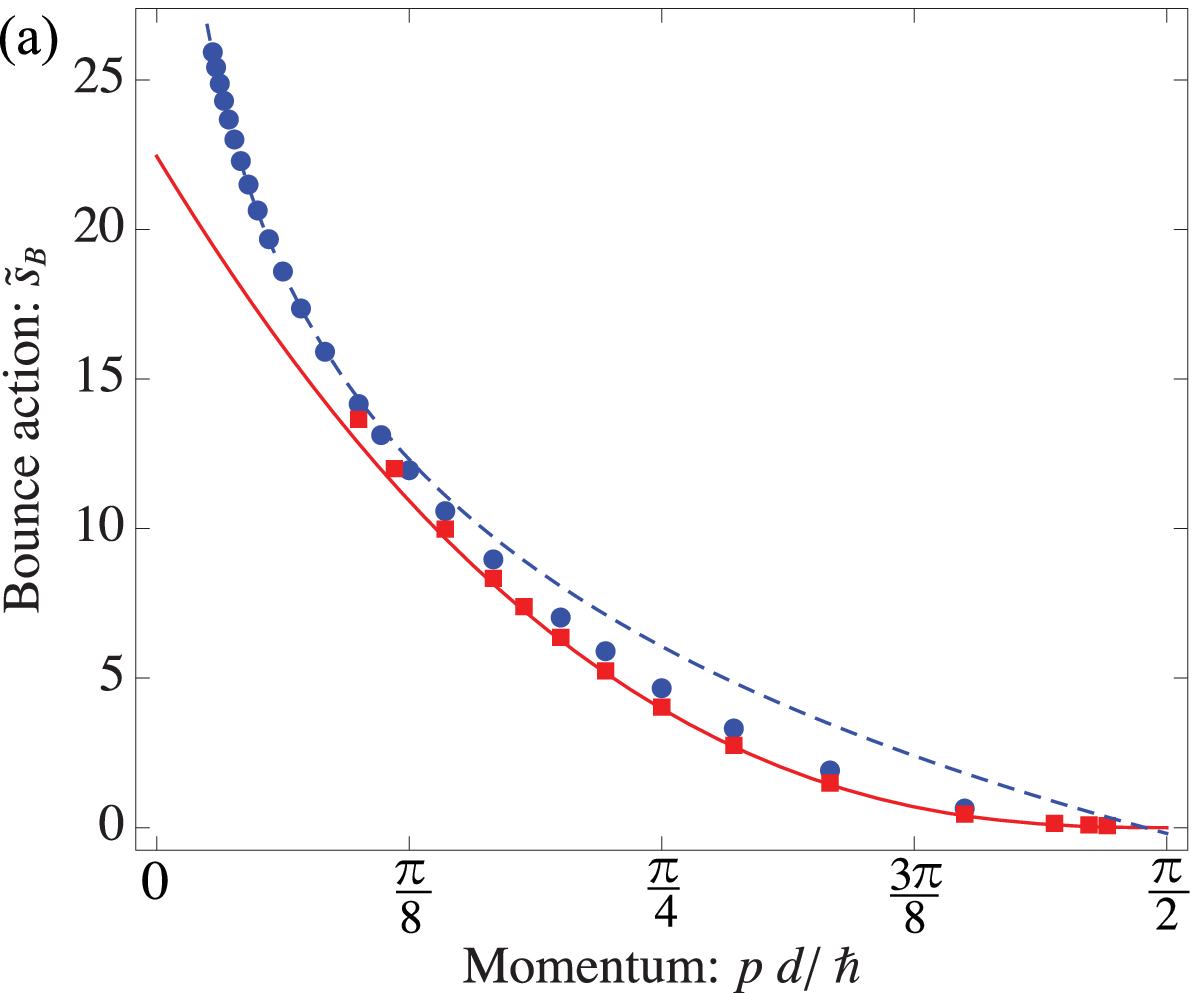}
\includegraphics[scale=0.6]{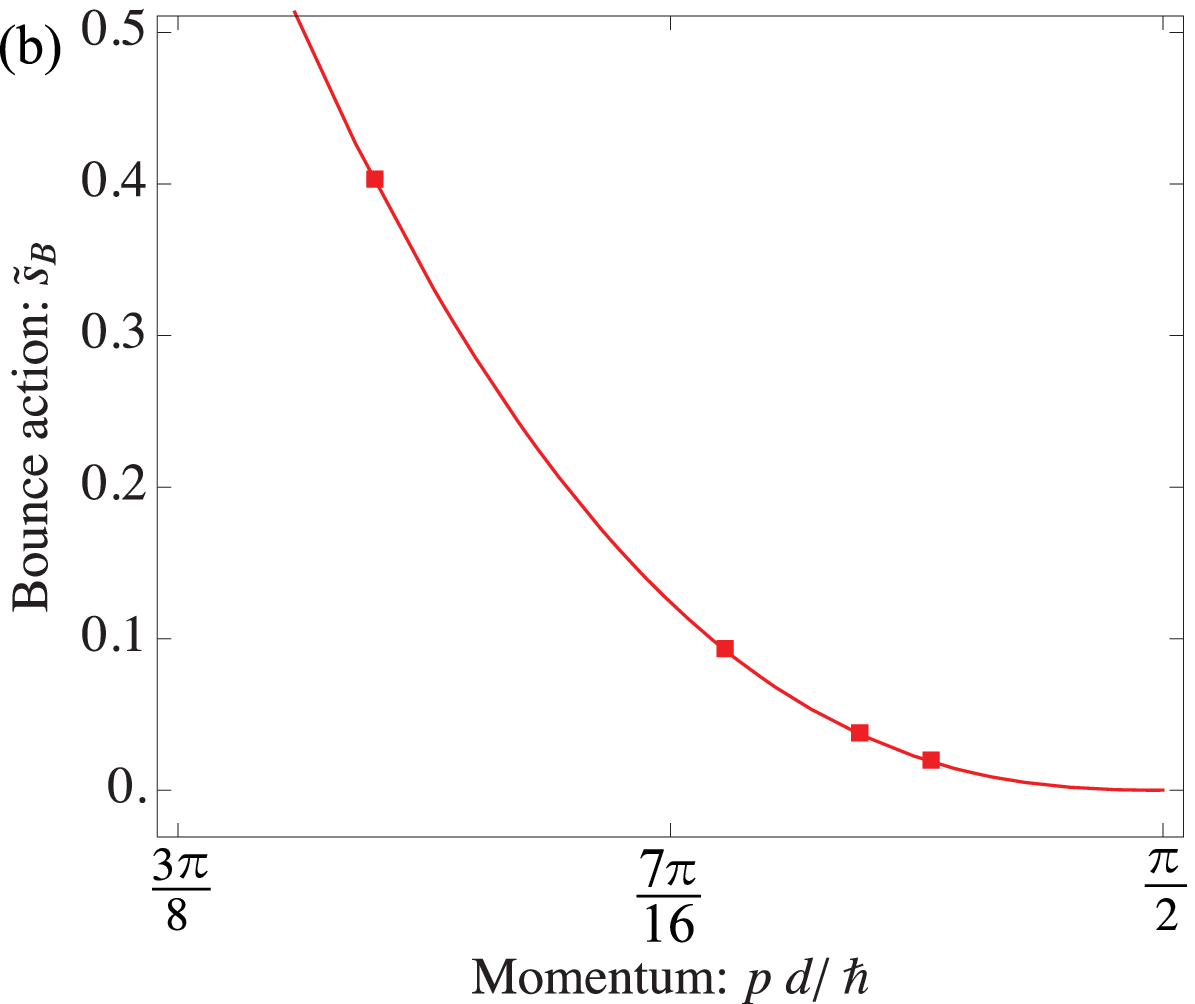}
\caption{\label{fig:SB}
(Color online) (a) The bounce action $\tilde{s}_B$ as a function of the (quasi-)momentum $p$.
The red squares represent $\tilde{s}_B$ obtained by the extrapolation shown in Fig.~\ref{fig:ConvSBA}, while the blue circles represent $\tilde{s}_B$ for the system size $L=2\pi\hbar /(pd)$, where the winding number is $n=1$. The red solid line represents $\tilde{s}_B=7.26(\pi/2-pd/\hbar)^{5/2}$. The blue dashed line represents the best fit to the data for $pd/\hbar \leq \pi/18$ with a function $f(x)=a\log(x) + b$, where the fitting parameters turn out to be $a=-9.04\pm 0.05$ and $b=3.84\pm 0.04$.
(b) Magnified view of (a) that focuses on the region near the critical momentum $pd/\hbar= \pi/2$.
}
\end{figure}
\begin{figure}[htb]
\includegraphics[scale=0.6]{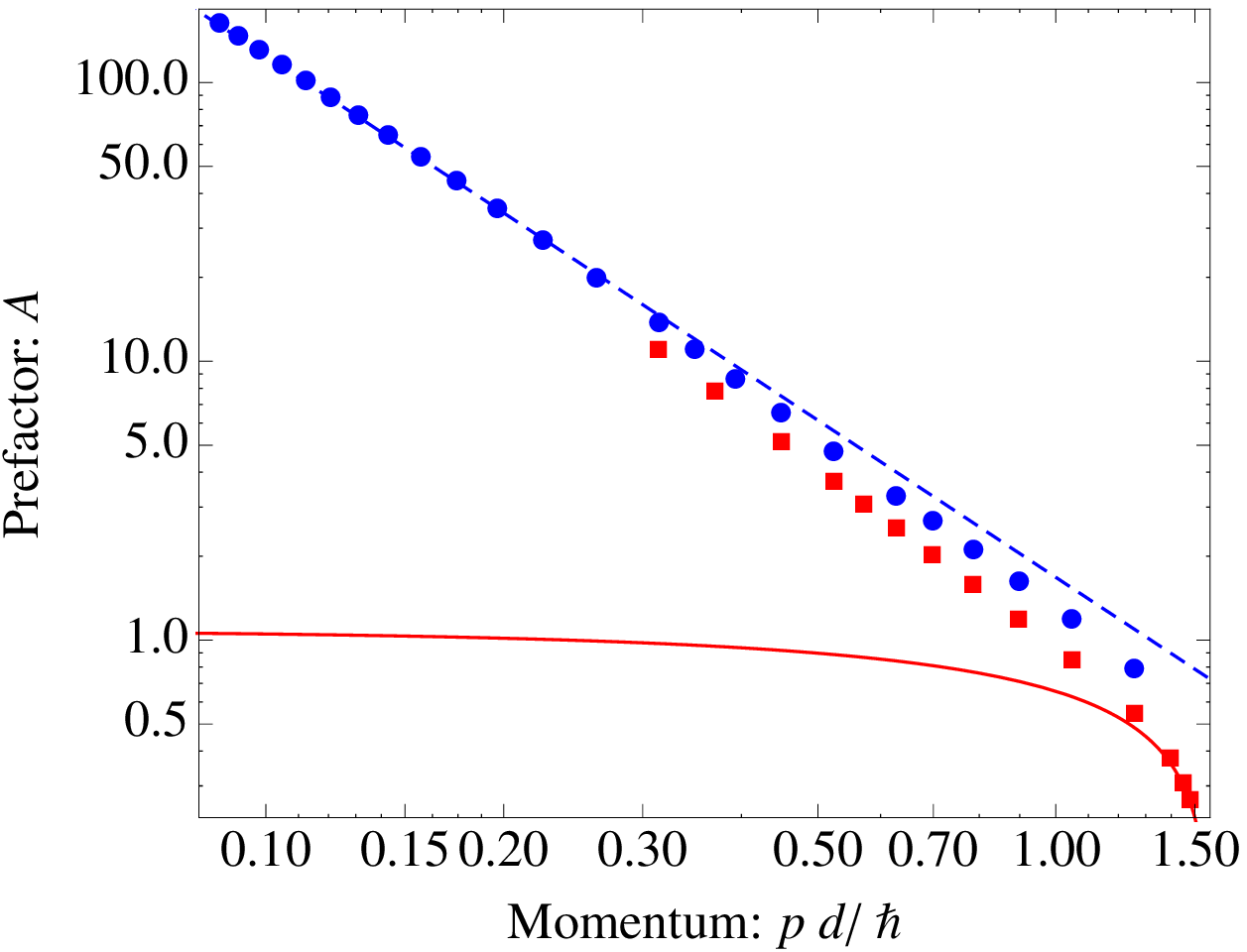}
\caption{\label{fig:AB}
(Color online) The prefactor $A$ as a function of the (quasi-)momentum $p$ is plotted in a log-log scale.
The red squares represent $A$ obtained by the extrapolation shown in Fig.~\ref{fig:ConvSBA}, while the blue circles represent $A$ for the system size $L=2\pi\hbar /(pd)$, where the winding number is $n=1$. The red solid line represents $A=0.867(\pi/2-pd/\hbar)^{1/2}$. The blue dashed line represents the best fit to the data for $pd/\hbar \leq \pi/18$ with a fitting function $f(x)=ax^{b}$, where the fitting parameters turn out to be $(a, b) = (1.68\pm 0.01, -1.87\pm 0.01)$.
}
\end{figure}

To obtain the bounce action $\tilde{s}_B$ and the prefactor $A$, we numerically calculate the bounce solution $\vec{\theta}^B(\tilde{\tau})$. The bounce solution $\vec{\theta}^B(\tilde{\tau})$ starts with the metastable state with the winding number $n$, i.e. $\vec{\theta}^B(-\infty) = \vec{\theta}_{n}^M$, goes through the saddle point $\vec{\theta}_{n}^S$, bounces at the classical turning point, and returns to the initial state. An example of the bounce solution is shown in Fig.~\ref{fig:bounce} for $n=1$ and $L=60$, where the origin of the time is set such that the bounce solution reaches the classical turning point at $\tilde{\tau}=0$. In Fig.~\ref{fig:bounce}(a), we see that the bounce solution for the quantum phase slips forms a pair of a vortex and an anti-vortex in the $(x,\tau)$-plane. As clearly seen in Fig.~\ref{fig:bounce}(b), the phase kink is not located at a lattice site, but at a link between two sites (30th and 31st), and thereby the density $n_j$ at any sites does not vanish in contrast to the phase-slips in continuous systems. This is consistent with the basic assumption of the quantum rotor model that the density fluctuation is small compared to the density itself.

Substituting the bounce solution into Eq.~(\ref{eq:sVec}), we obtain the bounce action. Moreover, solving the eigenvalue equations Eqs.~(\ref{eq:eigEqM1}) and (\ref{eq:eigEqM2}) with the bounce solution, we obtain $\vec{\xi}_m$ and $\vec{\xi}_m^{(0)}$. Once $\vec{\xi}_m$ and $\vec{\xi}_m^{(0)}$ are obtained, we can calculate the prefactor $A$ through Eq.~(\ref{eq:A}). 

When calculating the bounce solution, one often introduces collective variables to reduce the number of degrees of freedom~\cite{buechler-01, khlebnikov-05, anatoli-05}. However, the use of such collective variables restricts the analyses to a small region with respect to the momentum. In contrast, we deal with all the phase degrees of freedom, and this unbiased treatment enables us to more accurately obtain the nucleation rate for the entire region of the momentum.

Let us calculate the bounce action $\tilde{s}_{\rm B}$ and the prefactor $A$ as functions of the momentum $p$. For a given value of $p$,  $\tilde{s}_{\rm B}$ and $A$ depend on the number of lattice sites $L$. A typical example is shown in Fig.~\ref{fig:ConvSBA}, where  $\tilde{s}_{\rm B}$ and $A$ at $pd/\hbar = 2\pi/5$ are plotted by varying $L$. When $L$ increases, these two quantities are converged to certain asymptotic values. We extract the asymptotic values by fitting the numerical data to a function $f(x) = a+b x^{-c}$ as represented by the blue lines in Fig.~\ref{fig:ConvSBA}. This way allows us to obtain  the values of $\tilde{s}_{\rm B}$ and $A$ for the thermodynamic limit ($L\rightarrow \infty$). In Figs.~\ref{fig:SB} and \ref{fig:AB}, the bounce action $\tilde{s}_{\rm B}$ and the prefactor $A$ versus the momentum $p$ for the thermodynamic limit  are plotted by the red squares.

While the extrapolation to the thermodynamic limit is applicable for a region of relatively large momenta ($pd/\hbar \geq \pi/10$ in Figs.~\ref{fig:SB} and \ref{fig:AB}), it is practically difficult for very small momenta because the calculations for very large systems are required. For this reason, in the region of small momenta we calculate $\tilde{s}_B$ and $A$ only by taking the system size $L=2\pi\hbar/(pd)$, i.e. the winding number $n=1$. For instance, we take $L=40$ for $pd/\hbar = \pi/20$. In Figs.~\ref{fig:SB} and \ref{fig:AB}, $\tilde{s}_B$ and $A$ for $n=1$ are plotted by the blue circles. Although the values of $\tilde{s}_B$ and $A$ are a little overestimated without the extrapolation, the system size for a small momentum is so large that the deviation from the values in the thermodynamic limit is fairly small as already seen in the data points at $pd/\hbar = \pi/10$ ($L=20$) in Figs.~\ref{fig:SB} and \ref{fig:AB}.

In the vicinity of the critical momentum $p_c=\hbar\pi/(2d)$, it has been predicted by previous work~\cite{anatoli-05} that the bounce action scales as $\tilde{s}_B = C_s(\pi/2-pd/\hbar)^{5/2}$. In a similar way, the scaling formula for the prefactor can be obtained as $A=C_A(\pi/2-pd/\hbar)^{1/2}$. Inserting the numerical data of $\tilde{s}_{\rm B}$ and $A$ for $pd/\hbar =8\pi/17$ into these formulae, we obtain the coefficients $C_s=7.26$ and $C_A=0.867$.  As shown in Figs.~\ref{fig:SB} and \ref{fig:AB}, the formulae with these values of the coefficients agree well with the numerical data (red squares) for the momentum close to $p_c$. Surprisingly, we find that the agreement in $\tilde{s}_{\rm B}$ is almost perfect until $pd/\hbar \simeq \pi/6$ which is far away from $p_c$. Note that while the previous work has predicted $C_s = 7.1$ that is indeed close to our prediction, it is a little less accurate than ours because of the use of a variational ansatz~\cite{anatoli-05}.

For small momenta, $p\ll \hbar/d$, we numerically find that the bounce action exhibits a logarithmic dependence as 
\begin{eqnarray}
\tilde{s}_{\rm B} = a_s \log (pd/\hbar) + b_s,
\label{eq:sBscale}
\end{eqnarray}
and that the prefactor obeys a power law as 
\begin{eqnarray}
A = a_A (pd/\hbar)^{b_A}.
\label{eq:Ascale}
\end{eqnarray}
We extract the coefficients $a_s=-9.04$, $b_s=3.84$, $a_A=1.68$, and $b_A=-1.87$ by fitting these formulae to the numerical data of $\tilde{s}_{\rm B}$ and $A$ for $n=1$ represented by blue circles in Figs.~\ref{fig:SB} and \ref{fig:AB}. Since the formula is valid for small momenta, we used the data in the region of $pd/\hbar \leq \pi/18$ for the fitting.
Substituting Eqs~(\ref{eq:sBscale}) and (\ref{eq:Ascale}) into Eq.~(\ref{eq:decayGam}), the scaling formula for the decay rate is derived as $\hbar\Gamma / (LE_{\rm J}) \propto \left(pd/\hbar\right)^{2.03K - 1.87}$. From this numerical result, we argue that the nucleation rate obeys the following power law,
\begin{eqnarray}
\frac{\hbar\Gamma}{LE_J} = C_{\Gamma} \left(\frac{pd}{\hbar}\right)^{2K - 2},
\label{eq:GamPL}
\end{eqnarray}
where the coefficient $C_{\Gamma}$ is independent of the momentum $p$. The same scaling formula has been derived for 1D homogeneous superconductor with energy dissipation at the phase-slip centers~\cite{zaikin-97}. We emphasize that this agreement is not trivial because the model used in Ref.~\onlinecite{zaikin-97} is qualitatively different from ours in the sense that our model explicitly includes the lattice and does not include energy dissipation. As explained in Sec.~\ref{sec:dga}, this agreement is rooted in the fact that both models exhibit the Berezinskii-Kosteritz-Thouless (BKT) transition at $K=2$.

Since the resistance $R$ is related to the nucleation rate as $R\propto \Gamma /p$~\cite{langer-67}, Eq.~(\ref{eq:GamPL}) indicates that a 1D Bose fluid can flow with almost no resistance in a periodic potential as long as the flow velocity is sufficiently small. This is consistent with the previous numerical result that the dipole oscillations of 1D lattice bosons confined in a parabolic potential are hardly damped when the flow velocity is more than one order of magnitude smaller than the mean-field critical velocity~\cite{danshita-09-1}.

Note that in deriving Eq.~(\ref{eq:GamPL}) we ignored the logarithmic correction stemming from the factor $\sqrt{\tilde{s}_{\rm B}}$. The formula of Eq.~(\ref{eq:GamPL}) has been found through the numerical analyses on the discrete quantum rotor model Eq.~(\ref{eq:rotorS}). However, it is very likely that this formula is generally valid for the spinless TL liquid in a periodic potential because it is a low-energy property.

\begin{figure}[htb]
\includegraphics[scale=0.6]{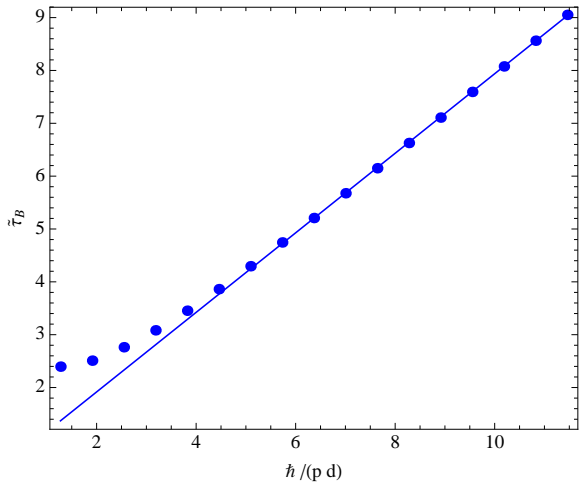}
\caption{\label{fig:VPdist}
The size of a vortex-antivortex pair $\tilde{\tau}_{\rm B}$ versus the inverse (quasi-)momentum $\hbar/(pd)$.  The blue line represents the best fit to the data for $pd/\hbar \leq \pi/18$ with a function $f(x)=ax+b$, where the fitting parameters turn out to be $a=0.753$ and $b=0.412$. Notice that the definition of $\tilde{\tau}_{\rm B}$ is depicted in Fig.~\ref{fig:bounce}(b). 
}
\end{figure}
\section{Qualitative discussions on the scaling formula}
\label{sec:dga}
In this section, we qualitatively explain a reason why the decay rate should obey the scaling formula of Eq.~(\ref{eq:GamPL}). Our explanation is twofold. First, the term $2K$ in the exponent $2K-2$ can be understood as the contribution from the bounce action, whose analytical expression can be obtained by using the analogy with the classical 2D XY model. Secondly, the term $-2$ in the exponent can be determined by considering the relation between the quantum nucleation of phase slips and the quantum phase transition between the superfluid and the Mott insulator.

Let us explain the first item. It is well known that taking the continuum limit and regarding $u\tau$ as another spatial variable $y$, the 1D quantum rotor action of Eq.~(\ref{eq:rotorS}) is equivalent to the energy of the classical 2D XY model.  In this mapping, the bounce solution corresponds to a vortex-antivortex pair. This analogy allows one to express the bounce action in terms of the distance between the vortex and antivortex $\tilde{\tau}_{\rm B}$. When $\tilde{\tau}_{\rm B}$ is sufficiently large compared to the size of vortex cores, the bounce action is well approximated as $S_{\rm B}/\hbar =2K\log\tilde{\tau}_{\rm B} + s_{\rm v} $, where $s_{\rm v}$ is the contribution from the vortex cores and independent of $\tilde{\tau}_{\rm B}$~\cite{yamada-95}. In order to find the $p$-dependence of $S_{\rm B}$, we numerically calculate $\tilde{\tau}_{\rm B}$ versus the inverse momentum $\hbar/(pd)$ for the winding number $n=1$ as shown in Fig.~\ref{fig:VPdist}. There we observe that when $pd/\hbar \ll 1$, $\tilde{\tau}_{\rm B}$ linearly increases with $\hbar/(pd)$. Using this relation, we obtain the $p$-dependence of the bounce action as $S_{\rm B}/\hbar = - 2K\log (pd/\hbar) + {\rm Const}$, and thereby $e^{-S_{\rm B}/\hbar}\propto (pd/\hbar)^{2K}$. Hence, it is natural to anticipate $\Gamma \propto L(pd/\hbar)^{2K+X}$, where $X$ is a constant that will be determined to be $X=-2$ in the following.

In order to corroborate $X=-2$, we focus on the nucleation rate from the state with a certain fixed winding number $n$, which is given by $\Gamma_{n \rightarrow n-1}\propto L^{-2K-X+1}$. We discuss the relation between the nucleation rate and the quantum superfluid-insulator transition. It is important to remind us that the instanton formula is derived within the dilute gas approximation (DGA), in which the bounces are assumed to be well separated from each other~\cite{coleman-77,callan-77,polyakov-77,coleman-85,sakita-85}. This means that when DGA breaks down, many vortices are created in the space-time coordinate so that they destroy the long-range phase coherence, leading to the quantum phase transition to an insulating phase. In other words, the breakdown of DGA signals the Mott transition~\cite{danshita-11}. In general, DGA is valid when the size of a bounce in the imaginary-time axis $\tau_{\rm B}$ is much smaller than the nucleation time $1/\Gamma$~\cite{coleman-85, sakita-85}. In the present case, the bounce time is inversely proportional to the momentum as shown in Fig.~\ref{fig:VPdist}, which means $\tau_{\rm B} \propto L$, and thereby $\tau_{\rm B}\Gamma_{n\rightarrow n-1} \propto L^{-2K-X+2}$. This means that when $-2K-X+2<0$, DGA is valid in the thermodynamic limit, i.e. the system is in the superfluid phase. Therefore, the condition $-2K-X+2=0$ has to be fulfilled at the Mott transition point. From a different point of view, the renormalization group analyses have shown that the transition of the BKT type occurs at $K=2$ in the presence of a periodic potential~\cite{giamarchi-04, cazalilla-11}. Thus, we reach the conclusion that $X=-2$. Notice that $X=-2$ can be derived in the same way also in the model of diffusive 1D superconductors studied in Ref.~\onlinecite{zaikin-97} from the fact that there exists the BKT transition at $K=2$ as well.

The explanation based on the relation between DGA and the superfluid-insulator transition is applicable also to the cases of a disorder potential and a single barrier. For the disorder case, Khlebnikov and Pryadko have found that the nucleation rate scales as $\Gamma \propto L (p/p_{\rm c})^{2K-1}$. Anticipating $\Gamma \propto L (p/p_{\rm c})^{2K+X}$, let us corroborate that $X=-1$ along the procedure introduced above. Since $\tau_{\rm B}\propto L$ as in the case of a periodic potential, the onset of the DGA breakdown, i.e. the transition point to an insulating phase, is given by $-2K-X+2=0$. Meanwhile, it is known from renormalization group analyses that $K=3/2$ at the transition to the Bose glass phase~\cite{giamarchi-88}. Therefore, a simple algebra leads to $X=-1$.  For the single-barrier case, it has been shown in Refs.~\onlinecite{kagan-00, buechler-01} that $\Gamma \propto (p/p_{\rm c})^{2K-1}$. Notice the absence of the factor of $L$, which reflects the fact that the phase slips occur only at the barrier potential. Anticipating $\Gamma \propto (p/p_{\rm c})^{2K+X}$, one can easily show that $X=-1$ as follows. The onset of the DGA breakdown is given by $-2K-X+1=0$ while renormalization group analyses have shown that a transition to an insulating phase pinned by the barrier occurs at $K=1$~\cite{kane-92}. Hence, $X=-1$. 

Furthermore, we briefly discuss the effects of finite temperatures on our results. For the scaling formula of Eq.~(\ref{eq:GamPL}) to be valid, the bounce time $\tau_{\rm B}$ has to be sufficiently small compared to the inverse temperature $\hbar/(k_{\rm B} T)$. Since $\tilde{\tau}_{\rm B} \sim \hbar/(pd)$ as shown in Fig.~\ref{fig:VPdist}, this condition turns out to be $k_{\rm B}T \ll E_{\rm J} pd/\hbar$. In the intermediate temperature region, namely $E_{\rm J} pd/\hbar \ll k_{\rm B}T \ll E_{\rm J}$, the bounce time is bounded by the inverse temperature as $\tilde{\tau}_{\rm B}\sim E_{\rm J}/(k_{\rm B}T)$. In this case, the nucleation of phase slips occurs due to the thermally assisted quantum tunneling~\cite{weiss-08}, and the nucleation rate is given by 
\begin{eqnarray}
\Gamma \propto L{pd \over \hbar}\left(\frac{k_{\rm B}T}{E_{\rm J}}\right)^{2K-c},
\label{eq:GamFT}
\end{eqnarray}
where the constant $c$ has been determined to be $c=3$ in previous work~\cite{lobos-11}. Notice that scaling formulae similar to Eq.~(\ref{eq:GamFT}) have been found also in the cases of a disorder potential~\cite{khlebnikov-05} and a single-barrier potential~\cite{kagan-00, buechler-01}.  Equation (\ref{eq:GamFT}) means that the transport is ohmic, but that the resistance can be very small when $K\gg 1$. This ensures the presence of superfluidity in the practical sense even at small finite temperatures~\cite{kagan-00}. When the temperature is as high as $k_{\rm B}T\gg E_{\rm J}$, the phase slips can not be nucleated in the space-time coordinate and the thermal activation process becomes dominant to the superflow decay.

Since the Josephson plasma energy $E_{\rm J}$ separates the quantum-tunneling regime from the thermal-activation regime, it is important to provide an estimate of $E_{\rm J}$ for present typical experiments. Taking the experiment of Ref.~\onlinecite{fertig-05} for example, $E_{\rm J}/k_{\rm B} \sim 30{\rm nK}$ because the sound velocity is $u \simeq 2.1{\rm mm/s}$ and the lattice spacing is $d=405{\rm nm}$. Given the fact that recent experiments have lowered the temperature as low as ${\rm 5nK}$~\cite{trotzky-10,sugawa-11}, it is very likely that the regime of $k_{\rm B}T \ll E_J$ is experimentally accessible so that one can observe the superflow decay via quantum nucleation of phase slips.

\section{TEBD analyses of the Bose-Hubbard model}
\label{sec:TEBD}
\begin{figure}[tb]
\includegraphics[scale=0.55]{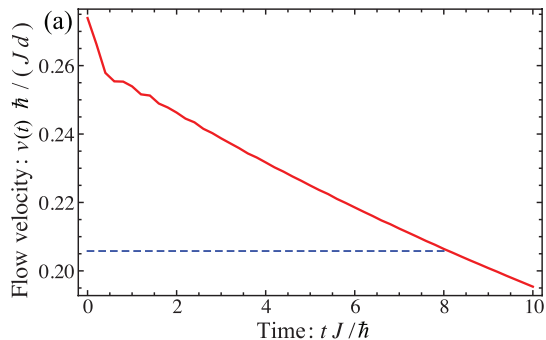}
\caption{\label{fig:velt}
(Color online) The red solid line represents the time evolution of the flow velocity $v(t)$ in the dynamics of the 1D BHM, where $L=160$, $\nu =1$, $U/J=3$, and $n=4$. The blue dashed line represents the flow velocity at the winding number $n=3$.
}
\end{figure}
\begin{figure}[tb]
\includegraphics[scale=0.55]{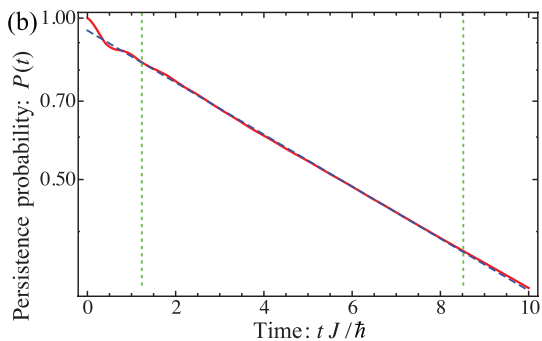}
\caption{\label{fig:Pt}
(Color online) The red solid line represents the time evolution of the persistence probability $P(t)$ in the dynamics of the 1D BHM, where $L=160$, $\nu = 1$, $U/J=3$, and $n=4$. The longitudinal axis is shown in a logarithmic scale. In the region sandwiched between the two green dotted lines, $P(t)$ decays exponentially. The blue dashed line represents the best fit with a function of Eq.~(\ref{eq:exfit}) to the data in the exponentially decaying region.
}
\end{figure}
While the scaling formula of Eq.~(\ref{eq:GamPL}) was obtained from the $O(2)$ quantum rotor model, it is expected to hold generally for 1D spinless superfluids in a periodic potential with an integer filling for the following two reasons. First, this formula is a property at small momenta ($p\ll p_{\rm c}$), and such a low-energy property should be valid commonly in the spinless TL liquid. For example, the compressibility and the long-range behaviors of correlation functions are expressed in terms of the TL parameter $K$ and the sound velocity $u$ regardless of microscopic details of the original Hamiltonian~\cite{giamarchi-04}. Secondly, as discussed in the previous section, this formula is closely related to the Mott transition at $K=2$, which is a common property in the spinless TL liquids with an integer filling in a periodic potential.

In this section, we study the superflow decay in the Bose-Hubbard model with unit filling in order to demonstrate that the applicability of the scaling formula is not limited to the quantum rotor model. Recall that the quantum rotor model quantitatively agrees with the Bose-Hubbard model only in the region of high filling factors ($\nu\gg 1$)~\cite{danshita-10}. It is important to investigate the unit-filling case also because the superfluid transport of 1D Bose gases in optical lattices has been experimentally studied mainly in the region of low-filling factors ($\nu \sim 1$).
In the low-filling regime, it has been predicted within the GP mean field theory that the Landau instability, which is characterized by the emergence of excitations with negative energies, sets in at momenta smaller than the critical value for the dynamical instability, i.e. $p = \hbar \pi /(2d)$~\cite{smerzi-02, wu-03}. However, this instability can break down superflow only when the temperature is sufficiently high so that the thermal fraction is comparable to the condensate fraction~\cite{sarlo-05, konabe-06, iigaya-06}, and is not relevant in the regime of our interest in which quantum fluctuations are dominant over thermal ones.

\begin{figure}[tb]
\includegraphics[scale=0.55]{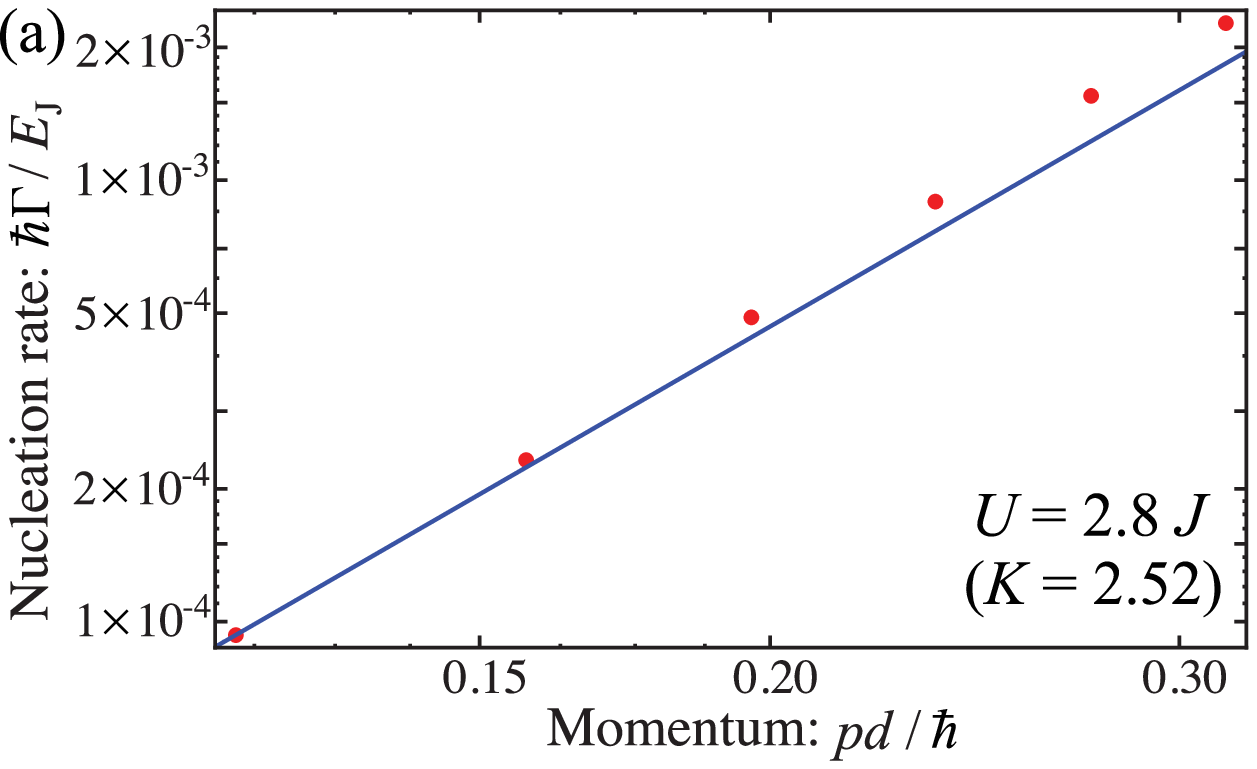}
\includegraphics[scale=0.55]{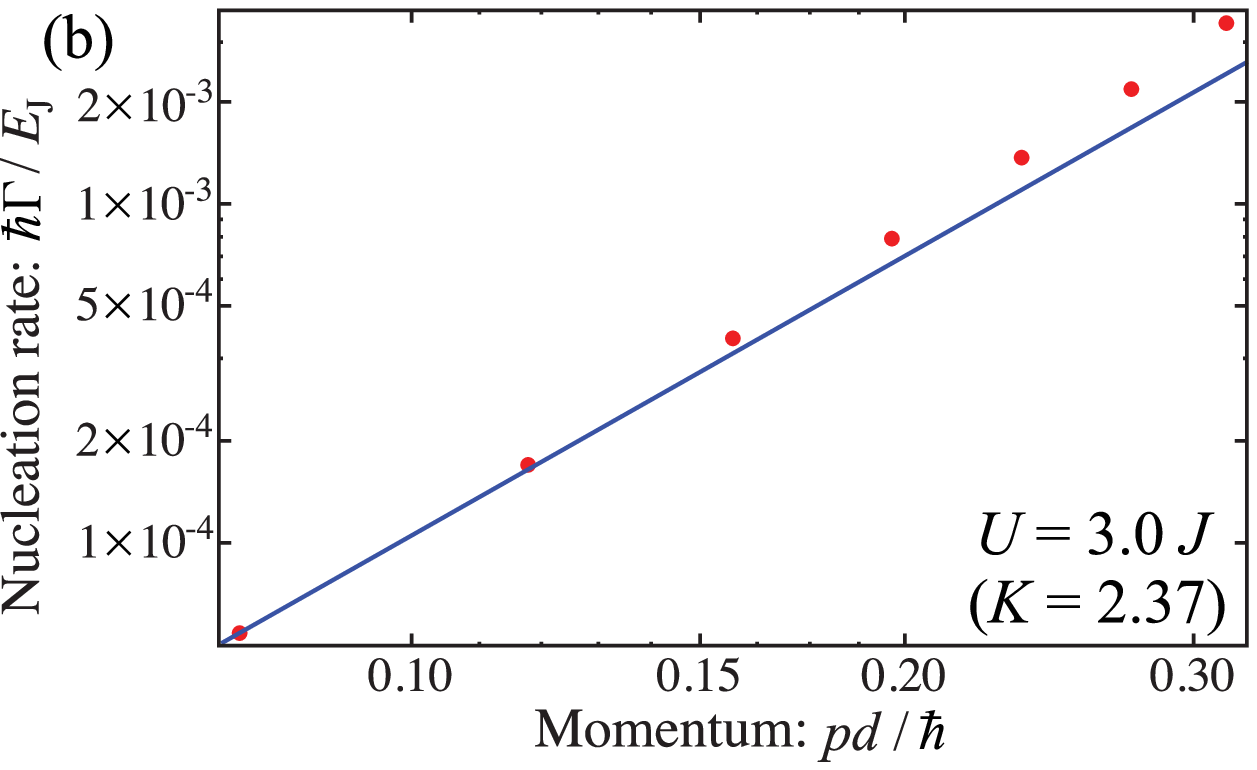}
\includegraphics[scale=0.55]{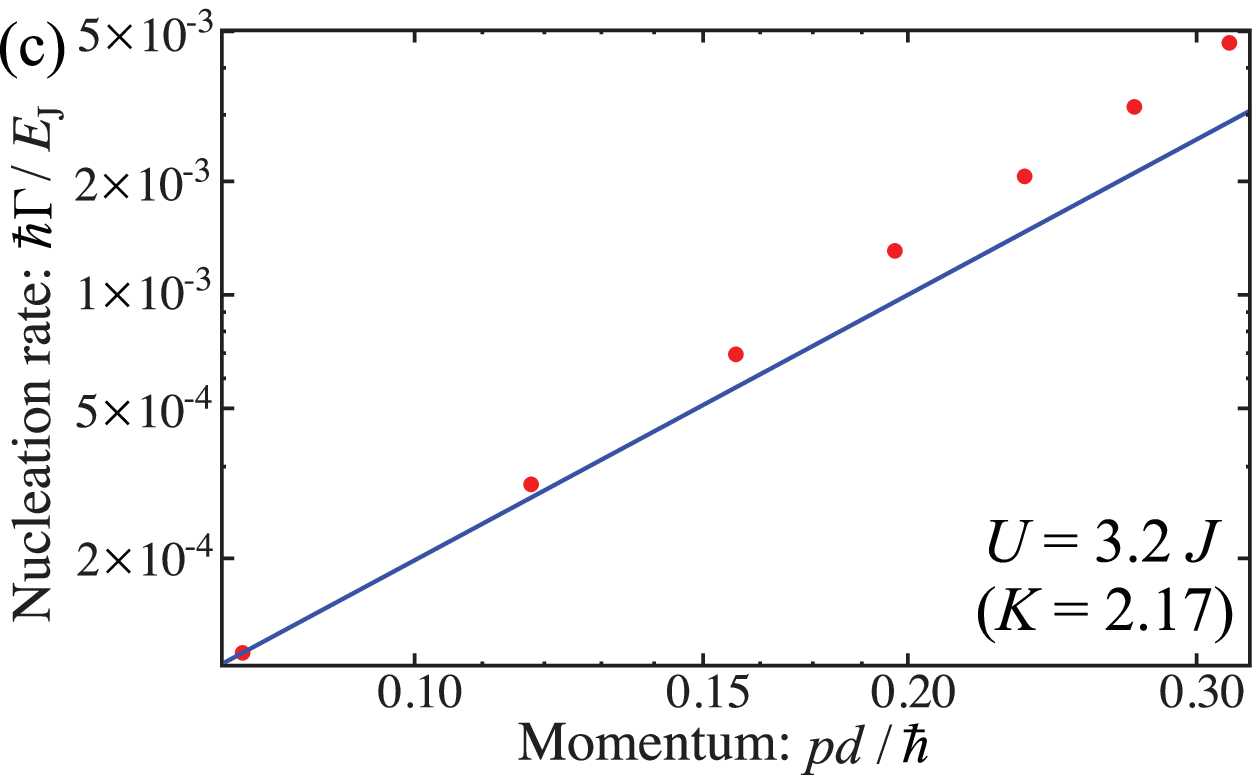}
\caption{\label{fig:NrateTEBD}
The red circles represent the nucleation rates of quantum phase slips $\Gamma$ extracted from the real-time dynamics of the 1D Bose-Hubbard model with $L=160$ as functions of the flow (quasi-)momentum $p$, where $U/J=2.8$ (a), $3$ (b), and $3.2$ (c). The plots are shown in a log-log scale. In each plot, the blue solid line represents the scaling formula of Eq.~(\ref{eq:GamPL}) with the constant $C_{\Gamma}$ determined such that the line passes on the data point with the smallest momentum. The TL parameters are taken from Ref.~\onlinecite{kuhner-00} as $K=2.52$ (a), $2.37$ (b), and $2.17$ (c).
}
\end{figure}

We describe 1D lattice bosons in a ring-shaped geometry with the following BHM with a phase twist:
\begin{eqnarray}
\hat{H} \!=-\! J\sum_{j=1}^L\left( e^{-i\theta} \hat{b}_j^{\dagger} \hat{b}_{j+1} + h.c. \right)
+\frac{U}{2} \sum_{j=1}^L \hat{n}_j(\hat{n}_j -1),
\label{eq:BHH1D}
\end{eqnarray}
where $\hat{b}_{L+1}\equiv \hat{b}_{1}$, reflecting the nature of the ring-shaped geometry. In Eq.~(\ref{eq:BHH1D}), we include the phase twist $e^{-i\theta}$ in the hopping term in order to control the winding number of states. To deal with the quantum dynamics of superflow of the 1D BHM Eq.~(\ref{eq:BHH1D}), we use the TEBD method~\cite{vidal-04} for a periodic boundary condition~\cite{danshita-09-2}, which allows us to accurately compute the time evolution of many-body wave functions in 1D quantum lattice systems. 
It has been shown in our previous work that TEBD is applicable to the problem of superflow dynamics associated with quantum phase slips~\cite{danshita-10}.
 We first calculate the ground state of Eq.~(\ref{eq:BHH1D}) with the phase twist $\theta = 2\pi n/L$ via the imaginary time evolution, and thereby a flowing state with the winding number $n$ that is metastable in the classical limit is prepared. Taking this state as the initial state and setting $\theta = 0$ at $t=0$, we compute the real-time evolution.   

In Fig.~\ref{fig:velt}, we show the time evolution of the averaged flow velocity,
\begin{eqnarray}
v=\frac{J d}{i\hbar N} \sum_j \langle \hat{b}_j^{\dagger}\hat{b}_{j+1} - h.c. \rangle,
\end{eqnarray}
for $L=160$, $U/J=3$, and $n=4$. We see that the flow velocity decreases in time, clearly exhibiting the superflow decay due to quantum tunneling. However, the averaged flow velocity does not exhibit a sudden drop by a quantized amount, which could be a characteristic of phase slips, but gradually decreases in time. This is because the phase slip jump is smoothened out by taking the quantum average of many events. In each event a phase slip occurs at a different time. Notice that the flow velocity is constant in time if one computes classical dynamics of the Gross-Pitaevskii equation neglecting quantum fluctuations. 

To quantify the tunneling rate from the metastable state, i.e. the nucleation rate of quantum phase slips, we calculate the overlap of the wave function with the initial state $P(t) =|\langle \Psi (t)|\Psi (0) \rangle|^2$, which can be interpreted as the persistence probability, i.e., the probability that the state remains in the initial state by the time $t$. It is expected from the tunneling theory that when the initial state is a metastable state, the persistence probability decays exponentially as $P(t) \simeq \exp(-\Gamma t)$ with the nucleation rate $\Gamma$~\cite{takagi-02}. Notice that the ability to calculate the persistence probability is a clear advantage of TEBD for a periodic boundary condition over TEBD for infinite systems that was used in Ref.~\onlinecite{schachenmayer-10}.

In Fig.~\ref{fig:Pt}, we show $P(t)$ with the same parameters as used in Fig.~\ref{fig:velt}. Indeed there is a large region where $P(t)$ exhibits the exponential decay. We extract the nucleation rate $\Gamma$ by fitting the data in the exponentially decaying region to the following function,
\begin{eqnarray}
g(t) = D \exp(-\Gamma t),
\label{eq:exfit}
\end{eqnarray}
and taking $D$ and $\Gamma$ as free parameters. In Fig.~\ref{fig:NrateTEBD}, we plot the nucleation rates versus the momentum $p$ for three values of $U/J$, where $L=160$. We also show the scaling formula of Eq.~(\ref{eq:GamPL}) represented by the (blue) solid lines, where $C_{\Gamma}$ is taken such that the line passes on the data point with the smallest momentum. We use the TL parameter $K$ numerically extracted from the single-particle correlation function $\langle \hat{b}_r^{\dagger} \hat{b}_0 \rangle$ with the distance $16\leq r \leq 32$ in Ref.~\onlinecite{kuhner-00}. Notice that the TL parameter $K$ in the present paper is equivalent to $1/K$ in the definition of Ref.~\onlinecite{kuhner-00}. We clearly see that the data points approach the lines when the momentum decreases, thus justifying the validity of the scaling formula for BHM with unit filling.

\vspace{5mm}
\section{conclusions}
\label{sec:sum}

In summary, we have studied the decay of superflow via quantum nucleation of phase slips in one-dimensional (1D) superfluids in the presence of a periodic potential. Within the quantum rotor regime, we used the instanton method to obtain the nucleation rate for all the region of the momentum $p$. When the momentum is close to the mean-field critical value $p_{\rm c}$, we improved the expression of nucleation rate that was previously obtained in Ref.~\onlinecite{anatoli-05}. For small momenta $p\ll p_{\rm c}$, we derived the scaling formula of the nucleation rate with respect to $p$, which is expressed in Eq.~(\ref{eq:GamPL}). We discussed the relation between the dilute gas approximation and the quantum superfluid-insulator transition in order to gain a unified physical interpretation of the scaling formulae for periodic, disorder, and single-barrier potentials. Applying the time-evolving block decimation method to the 1D Bose-Hubbard model with unit filling, we analyzed the quantum dynamics of the superflow decay and confirmed the validity of the scaling formula of Eq.~(\ref{eq:GamPL}).

While we have calculated the nucleation rate of quantum phase slips in order to characterize the superflow decay, it still remains ambiguous how the nucleation rate is related to the transport of 1D Bose gases in the presence of a trapping potential that has been studied in cold atom experiments~\cite{stoeferle-04, fertig-05, mun-07, haller-10}. Since the damping rate of the dipole oscillations has been often used to quantify the transport of trapped atomic gases, in our future work we will clarify direct connections of the nucleation rate with the damping rate.
\begin{acknowledgments}
The authors thank N. Prokof'ev, B. Svistunov, A. Tokuno and S. Tsuchiya for useful discussions and comments.
I.~D. thanks Boston University visitors program for hospitality. A. P. was supported by NSF DMR-0907039, AFOSR FA9550-10-1-0110, and the Sloan Foundation. The computation in this work was partially done using the RIKEN Cluster of Clusters facility.
\end{acknowledgments}


\end{document}